\documentclass[conference]{IEEEtran}
\IEEEoverridecommandlockouts
\usepackage{cite}
\usepackage{amsmath,amssymb,amsfonts}
\usepackage{algorithm}
\usepackage{algpseudocode}
\usepackage{graphicx}
\usepackage{textcomp}
\usepackage{xcolor}
\usepackage{url}
\usepackage{tikz}
\usepackage{balance}
\usetikzlibrary{arrows.meta,positioning,calc,fit,shapes.geometric,shapes.symbols}
\def\BibTeX{{\rm B\kern-.05em{\sc i\kern-.025em b}\kern-.08em
    T\kern-.1667em\lower.7ex\hbox{E}\kern-.125emX}}
\begin{document}

\title{nascTime: A Full-Stack 5G-TSN Bridge Simulation Framework with SDAP-Based QoS Mapping and IEEE 802.1AS Transparent Clock
\thanks{This publication has emanated from research conducted with the financial
support of Research Ireland under Grant number 13/RC/2077 P2.
For the purpose of Open Access, the author has applied a CC-BY public
copyright license to any Author Accepted Manuscript version arising from
this submission}
}

\author{\IEEEauthorblockN{Mohamed Seliem, Dirk Pesch, Utz Roedig, and Cormac Sreenan}
\IEEEauthorblockA{\textit{School of Computer Science and Information Technology} \\
\textit{National Universities of Ireland, University College Cork}, Cork, Ireland \\
MSeliem@ucc.ie, Dirk.pesch@ucc.ie, u.roedig@ucc.ie, Cormac.Sreenan@ucc.ie}
}

\maketitle

\begin{abstract}
3GPP Release~16 specifies how a 5G system can operate as a transparent IEEE~802.1 TSN bridge, yet no existing simulation framework implements the complete bridge architecture with end-to-end QoS mapping through the SDAP layer, per-flow Data Radio Bearer selection, and IEEE~802.1AS transparent clock behaviour with measured residence time. Existing tools model either QoS mapping without time synchronisation, or time synchronisation without a data plane.
This paper presents nascTime, a simulation framework built on OMNeT++~6.3, INET~4.6, and Simu5G that implements the full 3GPP 5G-TSN bridge model. The NW-TT and DS-TT are realised as modular compound modules that integrate with INET's \texttt{LayeredEthernetInterface} and streaming PHY. QoS mapping traverses the complete PCP\,$\rightarrow$\,DSCP\,$\rightarrow$\,QFI\,$\rightarrow$\,SDAP/DRB pipeline, and gPTP frames are transported through the simulated 5G radio path via L2-in-GTP-U encapsulation with per-message residence-time correction. We validate the framework with a three-endpoint factory topology under both ideal and fading channel conditions. In the ideal scenario, high-priority traffic achieves 99.9\% delivery with a mean end-to-end delay of 2.58\,ms, while the measured 5GS residence time exhibits a variance below 0.2\,$\mu$s. Under a fading channel, residence-time variance increases to 48\,$\mu$s, confirming that the framework captures radio-induced timing effects absent from abstract-delay simulators. nascTime is publicly available and constitutes the first full-stack 5G-TSN bridge simulation with SDAP-based QoS differentiation and measured IEEE~802.1AS transparent clock behaviour.
\end{abstract}


\section{Introduction}
IEEE 802.1 Time-Sensitive Networking (TSN) has become the predominant standard for deterministic industrial Ethernet~\cite{b1}, providing bounded latency, high reliability, and precise clock synchronisation through mechanisms such as IEEE~802.1Qbv time-aware shaping and IEEE~802.1AS generalised precision time synchronisation. Modern factory floors, however, increasingly require wireless connectivity for autonomous mobile robots, collaborative manipulators, and reconfigurable production cells, applications that wired TSN alone cannot serve.

5G New Radio with Ultra-Reliable Low-Latency Communication (URLLC) is well suited to fill this role. 3GPP Release~16 specifies how a 5G system operates as a logical TSN bridge, transparently connecting wired TSN segments through the wireless domain~\cite{b2}. The bridge architecture comprises a Network-side TSN Translator (NW-TT) at the User Plane Function, a Device-side TSN Translator (DS-TT) at the User Equipment, and a TSN Application Function (TSN~AF) that exposes bridge capabilities to the TSN Centralized Network Controller. The specification defines QoS mapping between TSN Priority Code Points (PCP) and 5G QoS Flow Identifiers (QFI), as well as IEEE~802.1AS transparent clock behaviour in which the 5G system reports its residence time in the gPTP correction field

Several simulation frameworks have been developed to study 5G-TSN integration. The 5GTQ framework~\cite{b3} implements NW-TT and DS-TT modules with QoS-aware priority scheduling, but maps traffic priority at the MAC scheduler level without utilizing the 3GPP SDAP layer for per-flow Data Radio Bearer (DRB) selection, and does not model gPTP time synchronization. Da Silva et al. \cite{b4} focus exclusively on time synchronization, modifying INET's IEEE 802.1AS model for UDP/IP transport over 5G, but do not implement data plane traffic forwarding through the bridge. The 6GDetCom Simulator \cite{b5} models TSN traffic shaping with a configurable wireless delay element, but does not simulate the actual 5G radio access network—replacing it with statistical delay distributions that cannot capture the impact of MAC scheduling, HARQ retransmissions, or multi-user resource contention.

In this paper, we present nascTime, a full-stack simulation framework for the 3GPP Release 16 5G-TSN bridge architecture. Our contributions are:
 
\begin{enumerate}
    \item Complete bridge architecture with SDAP-based QoS: nascTime implements the NW-TT and DS-TT as modular OMNeT++ compound modules that integrate with INET's \texttt{LayeredEthernetInterface}. The QoS pipeline maps TSN PCP through IPv4 DSCP to 5G QFI, which the SDAP layer~\cite{b18}—originally contributed by us to Simu5G v1.4.1 ~\cite{b19}—uses to select specific Data Radio Bearers, enabling per-flow scheduling at the gNB MAC layer.
    \item IEEE 802.1AS transparent clock with measured residence time: gPTP frames are transported through the actual simulated 5G radio path using L2-in-GTP-U encapsulation. A custom \texttt{GptpResidenceHeader} carries the NW-TT ingress timestamp, enabling the DS-TT to compute the actual 5G system residence time and update the IEEE 802.1AS correction field per message type.
    \item Multi-endpoint validation with bidirectional traffic: nascTime supports multiple UE/DS-TT pairs sharing a single gNB, with per-endpoint QoS configuration, gPTP replication, and bidirectional traffic. We validate the framework with a three-endpoint topology demonstrating near-perfect data delivery, correct time synchronization, and zero packet loss.
\end{enumerate}
 
The remainder of this paper is organized as follows. Section II provides background on the 5G-TSN bridge architecture and reviews related simulation frameworks. Section III describes the nascTime framework architecture. Section IV discusses implementation details and integration challenges. Section V presents validation results, and Section VI concludes the paper.

\section{Background and Related Work}
\subsection{3GPP 5G-TSN Bridge Architecture}
The 3GPP Release 16 specification TS 23.501 §5.28~\cite{b2} defines how a 5G system operates as one or more virtual IEEE 802.1 TSN bridges. As illustrated in Fig. 1, the bridge comprises two TSN Translator ports:

\textbf{The Network-side TSN Translator (NW-TT)}, located at the User Plane Function (UPF), connects the external TSN network to the 5G core. It translates between IEEE 802.1Q Ethernet frames and IP datagrams, mapping TSN QoS parameters to 5G QoS parameters for transport through the 5G system.

\textbf{The Device-side TSN Translator (DS-TT)}, located at the User Plane Function (UPF), connects the external TSN network to the 5G core. It translates between IEEE 802.1Q Ethernet frames and IP datagrams, mapping TSN QoS parameters to 5G QoS parameters for transport through the 5G system.

\textbf{The TSN Application Function (TSN AF)} provides the control plane interface between the 5G system and the TSN Centralized Network Controller (CNC). It exposes bridge capabilities including port-pair delays, supported traffic classes, and per-stream filtering and policing capabilities.

QoS mapping follows a multi-stage pipeline: the NW-TT reads the Priority Code Point (PCP) from the IEEE 802.1Q VLAN tag and maps it to an IPv4 Differentiated Services Code Point (DSCP). The UPF's Traffic Flow Filter maps DSCP to a 5G QoS Flow Identifier (QFI), which the SDAP layer uses to select a specific Data Radio Bearer (DRB). At the DS-TT, the process reverses: the IPv4 DSCP is mapped back to a PCP value in the reconstructed VLAN tag.

For time synchronization, the 5G system operates as an IEEE 802.1AS transparent clock. The NW-TT records the ingress timestamp when a gPTP frame enters the bridge. At the DS-TT, the residence time—the total delay through the 5G system, is computed and added to the gPTP correction field. This enables downstream TSN devices to account for the 5G bridge delay in their clock synchronization.

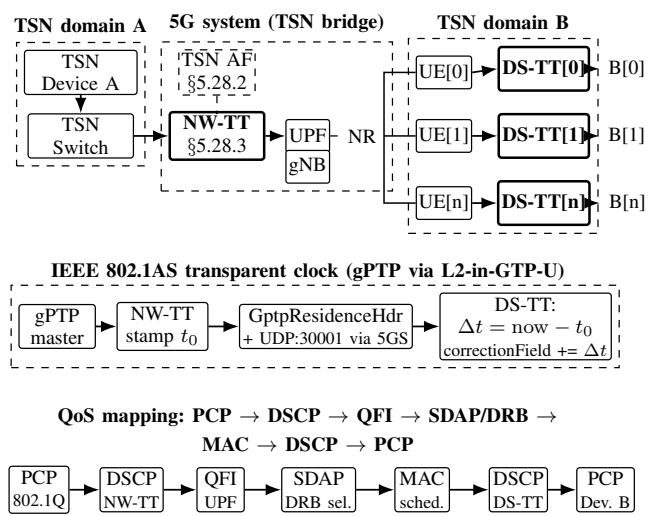
\begin{figure}[t]
\centering
\resizebox{\linewidth}{!}{
\begin{tikzpicture}[
    x=1in,y=1in,
    >=Latex,
    font=\footnotesize,
    line width=0.5pt,
    box/.style={
        draw,
        rounded corners=1.2pt,
        align=center,
        inner sep=1.5pt,
        minimum height=0.22in,
        fill=white
    },
    thickbox/.style={
        draw,
        rounded corners=1.2pt,
        line width=1pt,
        align=center,
        inner sep=1.5pt,
        minimum height=0.24in,
        fill=white
    },
    smallbox/.style={
        draw,
        rounded corners=1pt,
        align=center,
        inner sep=1pt,
        minimum height=0.18in,
        fill=white
    },
    group/.style={
        draw,
        dashed,
        inner sep=0.04in
    },
    arr/.style={->, line width=0.7pt},
    conn/.style={line width=0.6pt}
]

% =========================================================
% TOP ROW
% total width kept ~3.35in
% =========================================================

% --- Domain A
\node[box, minimum width=0.62in] (devA) at (0.45,-0.20) {TSN\\Device A};
\node[box, minimum width=0.58in] (swA)  at (0.45,-0.55) {TSN\\Switch};
\draw[arr] (devA) -- (swA);

\node[group, fit=(devA)(swA),
      label={[font=\footnotesize\bfseries]90:TSN domain A}] (gA) {};

% --- 5G bridge
\node[smallbox, dashed, minimum width=0.42in] (af) at (1.2,-0.2) {TSN AF\\$\S$5.28.2};
\node[thickbox, minimum width=0.52in] (nwtt) at (1.2,-0.55) {\textbf{NW-TT}\\$\S$5.28.3};
\node[smallbox, minimum width=0.24in] (upf) at (1.7,-0.55) {UPF};
\node[smallbox, minimum width=0.24in] (gnb) at (1.7,-0.72) {gNB};
\node[font=\footnotesize] (nr) at (2.0,-0.55) {NR};

\draw[dashed] (af.south) -- ++(0,-0.1) -| (nwtt.north);
\draw[arr] (swA.east) -- (nwtt.west);
\draw[arr] (nwtt.east) -- (upf.west);
\draw[dashed] (upf.east) -- (nr.west);

\node[group, fit=(af)(nwtt)(upf)(gnb)(nr),
      label={[font=\footnotesize\bfseries]90:5G system (TSN bridge)}] (g5g) {};

% --- Domain B
\node[smallbox, minimum width=0.22in] (ue0) at (2.45,-0.2) {UE[0]};
\node[smallbox, minimum width=0.22in] (ue1) at (2.45,-0.55) {UE[1]};
\node[smallbox, minimum width=0.22in] (uen) at (2.45,-0.92) {UE[n]};

\node[thickbox, minimum width=0.36in] (ds0) at (3.,-0.18) {\textbf{DS-TT[0]}};
\node[thickbox, minimum width=0.36in] (ds1) at (3.,-0.55) {\textbf{DS-TT[1]}};
\node[thickbox, minimum width=0.36in] (dsn) at (3.,-0.92) {\textbf{DS-TT[n]}};

\draw[arr] (ue0.east) -- (ds0.west);
\draw[arr] (ue1.east) -- (ds1.west);
\draw[arr] (uen.east) -- (dsn.west);

\draw[arr] (ds0.east) -- ++(0.05,0) node[right] {B[0]};
\draw[arr] (ds1.east) -- ++(0.05,0) node[right] {B[1]};
\draw[arr] (dsn.east) -- ++(0.05,0) node[right] {B[n]};

% NR branching
\coordinate (split) at (2.12,-0.55);
\draw[conn] (nr.east) -- (split);
\draw[conn] (split) |- (ue0.west);
\draw[conn] (split) -- (ue1.west);
\draw[conn] (split) |- (uen.west);

\node[group, fit=(ue0)(uen)(ds0)(dsn),
      label={[font=\footnotesize\bfseries]90:TSN domain B}] (gB) {};

% =========================================================
% MIDDLE BLOCK
% =========================================================

\node[font=\footnotesize\bfseries] at (1.70,-1.3)
{IEEE 802.1AS transparent clock (gPTP via L2-in-GTP-U)};

\node[box, minimum width=0.42in] (gm)    at (0.32,-1.6) {gPTP\\master};
\node[box, minimum width=0.50in] (stamp) at (0.9,-1.6) {NW-TT\\stamp $t_0$};
\node[box, minimum width=0.64in] (hdr)   at (1.8,-1.6) {GptpResidenceHdr\\{\scriptsize + UDP:30001 via 5GS}};
\node[box, minimum width=0.78in] (corr)  at (2.90,-1.6) {DS-TT: \\$\Delta t=\mathrm{now}-t_0$\\{\scriptsize correctionField += $\Delta t$}};

\draw[arr] (gm) -- (stamp);
\draw[arr] (stamp) -- (hdr);
\draw[arr] (hdr) -- (corr);

\node[group, fit=(gm)(stamp)(hdr)(corr)] (gMid) {};

% =========================================================
% BOTTOM BLOCK
% =========================================================

\node[font=\footnotesize\bfseries] at (1.70,-2.1)
{QoS mapping: PCP $\rightarrow$ DSCP $\rightarrow$ QFI $\rightarrow$ SDAP/DRB $\rightarrow$ };

\node[font=\footnotesize\bfseries] at (1.70,-2.25)
{MAC $\rightarrow$ DSCP $\rightarrow$ PCP};

\node[box, minimum width=0.28in] (pcp1)  at (0.22,-2.5) {PCP\\{\scriptsize 802.1Q}};
\node[box, minimum width=0.32in] (dscp1) at (0.73,-2.5) {DSCP\\{\scriptsize NW-TT}};
\node[box, minimum width=0.26in] (qfi)   at (1.22,-2.5) {QFI\\{\scriptsize UPF}};
\node[box, minimum width=0.34in] (sdap)  at (1.76,-2.5) {SDAP\\{\scriptsize DRB sel.}};
\node[box, minimum width=0.30in] (mac)   at (2.33,-2.5) {MAC\\{\scriptsize sched.}};
\node[box, minimum width=0.32in] (dscp2) at (2.86,-2.5) {DSCP\\{\scriptsize DS-TT}};
\node[box, minimum width=0.28in] (pcp2)  at (3.34,-2.5) {PCP\\{\scriptsize Dev. B}};

\draw[arr] (pcp1) -- (dscp1);
\draw[arr] (dscp1) -- (qfi);
\draw[arr] (qfi) -- (sdap);
\draw[arr] (sdap) -- (mac);
\draw[arr] (mac) -- (dscp2);
\draw[arr] (dscp2) -- (pcp2);

\end{tikzpicture}
}
\caption{3GPP 5G-TSN bridge architecture with multi-endpoint topology and IEEE 802.1AS transparent clock via L2-in-GTP-U.}
\label{fig:tsn-5g-bridge}
\end{figure}

\subsection{Related Simulation Frameworks}
Table~\ref{tab:comparison_alt} compares nascTime with existing 5G-TSN simulation frameworks across key architectural features.

\begin{table*}[t]
\centering
\caption{Comparison of 5G-TSN Simulation Frameworks}
\label{tab:comparison_alt}
\renewcommand{\arraystretch}{1.2}
\begin{tabular}{|l|c|c|c|c|c|}
\hline
\textbf{Feature} & \textbf{5GTQ~\cite{b3}} & \textbf{da Silva~\cite{b4}} & \textbf{6GDetCom~\cite{b5}} & \textbf{Satka~\cite{b17}} & \textbf{nascTime}~\cite{b20} \\
\hline
\hline
5G Radio Stack & Simu5G & Simu5G & Abstract delay & NeSTiNg+delay & Simu5G \\
\hline
INET Version & 4.4.1 & 4.x & 4.5.2 & 4.x & 4.6 \\
\hline
NW-TT / DS-TT & $\checkmark$ & Sync only & $\times$ & $\checkmark$ & $\checkmark$ \\
\hline
QoS Mapping & Priority & $\times$ & TSN shaping & PCP$\leftrightarrow$priority & PCP$\rightarrow$QFI$\rightarrow$DRB \\
\hline
SDAP / DRB & $\times$ & $\times$ & $\times$ & $\times$ & $\checkmark$ \\
\hline
gPTP Transport & $\times$ & Mod. 802.1AS & Basic & $\times$ & L2-in-GTP-U \\
\hline
Residence Time & $\times$ & Implicit & Configured & $\times$ & Measured \\
\hline
Multi-Endpoint & $\times$ & $\times$ & $\checkmark$ (AGVs) & $\times$ & $\checkmark$ (3+ UEs) \\
\hline
Bidirectional & $\times$ & $\times$ & Partial & $\times$ & $\checkmark$ \\
\hline
Streaming PHY & -- & -- & -- & -- & $\checkmark$ \\
\hline
TSN AF / BMCA & $\times$ & $\times$ & $\times$ & CNC (WiP) & $\checkmark$ \\
\hline
\end{tabular}
\end{table*}

The 5GTQ framework~\cite{b3} was among the first to address QoS-aware 5G-TSN simulation, implementing NW-TT and DS-TT modules with priority-based scheduling. However, 5GTQ maps traffic priority directly at the MAC scheduler level using 5QI values without utilising the 3GPP SDAP layer for per-flow DRB selection. All traffic therefore shares a single radio bearer regardless of its TSN priority class. Furthermore, 5GTQ does not model gPTP time synchronisation or residence time measurement.

Da Silva et al.~\cite{b4} address time synchronization by modifying INET's IEEE 802.1AS model to operate over UDP/IP through the 5G network, demonstrating synchronization accuracy below one microsecond. Their work supports 3GPP Release 17 time-aware system modes and validates synchronization in multiple clock master configurations. However, their framework does not implement data plane traffic forwarding—only gPTP control frames are transported through the 5G system. Without NW-TT/DS-TT data forwarding, QoS mapping, or traffic delivery validation, the framework cannot evaluate the bridge's performance under realistic mixed traffic conditions.

The 6GDetCom Simulator~\cite{b5}, developed within the EU DETERMINISTIC6G project, models end-to-end deterministic communication across heterogeneous domains including 6G wireless segments. The wireless domain is represented as a configurable delay element with statistical characteristics (mean, variance, distribution), enabling analysis of TSN scheduling algorithms such as Time-Aware Shaping under wireless delay uncertainty. However, the simulator does not include an actual 5G radio access network---there is no MAC scheduler, HARQ, channel model, or resource block allocation. Consequently, it cannot capture the effects of multi-user contention, fading, or scheduling discipline on bridge performance.

Additional related work includes Muslim et al.~\cite{b6}, who extend the 802.1AS-over-UDP/IP approach with 5G handover scenarios; the P5G-TSN framework~\cite{b7}, which extends 5GTQ with TDD pattern analysis; and Wang et al.~\cite{b8}, who study time synchronisation collisions in large-scale 5G+TSN networks. None of these frameworks implement the complete QoS pipeline through the SDAP layer or validate multi-endpoint data delivery with measured residence time.

Satka et al. provide a systematic review of TSN-5G integration~\cite{b15}, identifying time synchronisation as the dominant research focus (73\% of studies) and QoS mapping as an open challenge. Their QoS-MAN algorithm~\cite{b16} maps TSN flows to 5G QoS flows based on deadline, jitter, bandwidth, and packet loss constraints, and they developed an early NeSTiNg-based translation technique~\cite{b17}. QoS-MAN operates at the flow-to-5QI assignment level and has not been evaluated within a full-stack simulation including the SDAP/DRB pipeline and measured residence time. nascTime provides the simulation infrastructure to evaluate such algorithms under realistic radio conditions.
 
No existing simulation framework combines (1) actual 5G radio simulation with MAC scheduling and channel models, (2) end-to-end QoS mapping through the SDAP/DRB pipeline, (3) measured IEEE 802.1AS residence time from actual radio path traversal, (4) multi-endpoint scaling with bidirectional traffic validation, and (5) integration with INET's \texttt{LayeredEthernetInterface} for TSN feature compatibility. nascTime addresses all five requirements.

\section{nascTime Framework Architecture}

\subsection{NW-TT Module}
\begin{figure}[t]
\centering
\resizebox{\linewidth}{!}{
\begin{tikzpicture}[
    x=1in,y=1in,
    >=Latex,
    font=\footnotesize,
    line width=0.5pt,
    mod/.style={
        draw,
        rounded corners=2pt,
        align=center,
        inner sep=2pt,
        minimum height=0.22in,
        text width=#1,
        fill=white
    },
    note/.style={
        draw,
        dashed,
        rounded corners=2pt,
        align=center,
        inner sep=2pt,
        minimum height=0.16in,
        text width=#1
    },
    arr/.style={->, line width=0.7pt},
    darr/.style={->, dashed, line width=0.7pt},
]

% Outer container
\draw[rounded corners=6pt, dashed] (0.05,0.05) rectangle (3.35,2.05);
\node[font=\footnotesize\bfseries] at (1.70,2.14)
{NW-TT compound module architecture};
\node[font=\footnotesize\bfseries] at (1.70,1.96)
{(extends NetworkLayerNodeBase)};

% Left stack
\node[draw, rounded corners=2pt, fill=blue!20, minimum width=0.78in, minimum height=0.30in, align=center] (ethif) at (0.65,0.45)
{ethIf\\{\scriptsize LayeredEthernet}};
\node[font=\scriptsize] at (0.65,0.18) {StreamingPhyLayer};

\node[draw, rounded corners=2pt, fill=violet!20, minimum width=0.78in, minimum height=0.22in, align=center] (ethli) at (0.65,0.92)
{ethLi (dispatcher)};

\node[draw, rounded corners=2pt, fill=orange!25, minimum width=0.78in, minimum height=0.22in, align=center] (encap) at (0.65,1.50)
{encap\\{\scriptsize EthernetEncapsulation}};

% Center stack
\node[draw, rounded corners=2pt, fill=green!20, minimum width=0.92in, minimum height=0.42in, align=center] (translator) at (2.1,1.5)
{\texttt{NwTtTranslator}\\{\scriptsize DSCP $\leftrightarrow$ PCP}\\{\scriptsize PCP-aware Ethernet path}};

\node[draw, rounded corners=2pt, fill=gray!15, minimum width=0.92in, minimum height=0.22in, align=center] (nl) at (2.1,0.92)
{network/link\\(dispatcher)};

\node[draw, dashed, rounded corners=2pt, minimum width=0.92in, minimum height=0.18in, align=center] (ipv4) at (2.1,0.55)
{{\scriptsize IPv4 routing (egress)}};

% Right block
\node[draw, rounded corners=2pt, fill=gray!15, minimum width=0.66in, minimum height=0.30in, align=center] (pppif) at (3.00,0.45)
{pppIf\\{\scriptsize \texttt{PppInterface}}};

% Internal arrows
\draw[arr] (ethif.north) -- node[left,font=\scriptsize]{ingress} (ethli.south);
\draw[arr] (ethli.north) -- node[left,font=\scriptsize]{egress} (encap.south);

\draw[arr] (encap.east) -- ++(0.14,0) |- node[pos=0.30,above,font=\scriptsize]{lower} (translator.west);
\draw[arr] (encap.east) -- ++(0.18,0) |- node[pos=0.75,above,font=\scriptsize]{upper} (nl.west);

\draw[arr] (translator.south) -- node[right,font=\scriptsize]{ipForwardOut} (nl.north);
\draw[arr] (nl.east) --  ++(0.37,0) |- node[pos=0.75,above,font=\scriptsize]{}(pppif.north);

\draw[darr] (nl.south) -- (ipv4.north);

% External ports
\node[font=\scriptsize, align=right] (tsnlab) at (-0.02,0.45) {TSN\\Switch};
\draw[arr] (tsnlab.east) -- (ethif.west);

\node[font=\scriptsize, align=left] (upflab) at (3.5,0.45) {UPF};
\draw[arr] (pppif.east) -- (upflab.west);

\end{tikzpicture}
}
\caption{NW-TT compound module architecture.}
\label{fig:nwtt_architecture}
\end{figure}
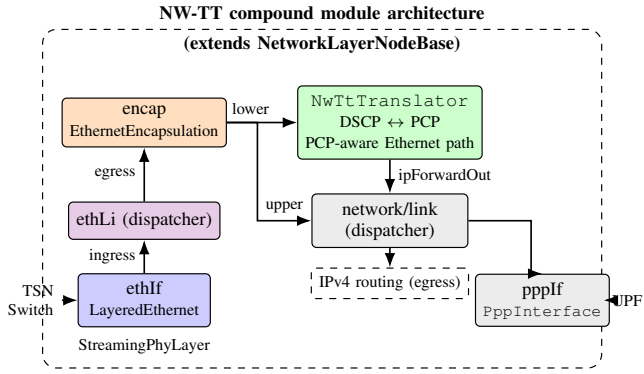

The NW-TT is implemented as a compound module extending INET's NetworkLayerNodeBase, following the same architectural pattern as Simu5G's UPF module. Fig.~\ref{fig:nwtt_architecture} shows the internal structure. The TSN-facing port uses INET's \texttt{LayeredEthernetInterface} with \texttt{EthernetStreamingPhyLayer}, connected through a \texttt{EthernetStreamingPhyLayer} (ethLi) to the \texttt{NwTtTranslator} simple module. The UPF-facing port uses a standard \texttt{PppInterface} connected through the inherited network layer dispatcher. On the ingress path (TSN → 5GS), the translator receives complete Ethernet frames from the TSN switch, detects gPTP frames by \texttt{ethertype} (0x88F7), and processes data and gPTP frames differently:
 
\textbf{Data frames:} The translator strips the IEEE 802.1Q VLAN tag, reads the PCP value, maps it to an IPv4 DSCP field, and forwards the IPv4 datagram to the UPF via the PPP interface. The UPF's Traffic Flow Filter reads the DSCP and sets the corresponding QFI in the GTP-U tunnel header.
 
\textbf{gPTP frames:} The translator prepends a custom \texttt{GptpResidenceHeader} carrying the current simulation time as the ingress timestamp, wraps the complete gPTP Ethernet frame in a UDP packet (port 30001), and sends it as an IPv4 datagram through the 5G system. For multi-endpoint topologies, the gPTP frame is replicated to all registered downstream devices.
 
On the egress path (5GS → TSN), return traffic from the UPF is routed by the NW-TT's IPv4 stack through an EthernetEncapsulation module, which adds MAC framing and sends the frame to the TSN switch via the ethLi dispatcher. Multi-endpoint support is achieved through a JSON-configurable array of \textit{{address, ue}} pairs. At initialization, the translator registers each downstream TSN device's IP address with the Simu5G binder, mapping it to the corresponding UE's NR node identifier obtained from the binder's node information map.

\subsection{DS-TT Module}
\begin{figure}[t]
\centering
\resizebox{\linewidth}{!}{
\begin{tikzpicture}[
    x=1in,y=1in,
    >=Latex,
    font=\footnotesize,
    line width=0.5pt,
    box/.style={
        draw,
        rounded corners=2pt,
        align=center,
        inner sep=2pt,
        minimum height=0.24in
    },
    bigbox/.style={
        draw,
        rounded corners=3pt,
        align=center,
        inner sep=2pt,
        minimum height=0.52in
    },
    arr/.style={->, line width=0.75pt},
    garr/.style={->, line width=0.75pt, draw=green!50!black},
    lab/.style={font=\scriptsize}
]

% -------------------------------------------------
% Outer frame
% -------------------------------------------------
\draw[dashed, rounded corners=6pt] (0.05,0.05) rectangle (3.35,1.95);

\node[font=\footnotesize\bfseries] at (1.70,2.08)
{DS-TT compound module architecture};

\node[font=\footnotesize\bfseries] at (1.70,1.84)
{(two-port L2 bridge)};

% -------------------------------------------------
% Bottom interfaces
% -------------------------------------------------
\node[bigbox, fill=blue!20, draw=blue!60!black, minimum width=0.86in]
    (ueeth) at (0.68,0.46) {ueEth\\{\scriptsize LayeredEthernet}};

\node[bigbox, fill=blue!20, draw=blue!60!black, minimum width=1.00in]
    (tsneth) at (2.66,0.46) {tsnEth\\{\scriptsize LayeredEthernet}};

\node[lab, align=center] at (0.68,0.08)
{EthernetPhyLayer\\(non-streaming)};

\node[lab, align=center] at (2.66,0.08)
{StreamingPhyLayer\\(cut-through)};

% external arrows
\node[lab, align=right] at (-0.02,0.48) {UE\\eth[0]};
\draw[arr] (0.03,0.48) -- (ueeth.west);

\node[lab, align=left] at (3.42,0.48) {TSN\\Device B};
\draw[arr] (tsneth.east) -- (3.30,0.48);

% -------------------------------------------------
% Dispatchers
% -------------------------------------------------
\node[box, fill=violet!20, draw=violet!60!black, minimum width=0.86in]
    (ueli) at (0.68,0.95) {ueLi (dispatcher)};

\node[box, fill=violet!20, draw=violet!60!black, minimum width=1.00in]
    (tsnli) at (2.66,0.95) {tsnLi (dispatcher)};

\draw[arr] (ueeth.north) -- (ueli.south);
\draw[arr] (tsneth.north) -- (tsnli.south);

% -------------------------------------------------
% Translator
% -------------------------------------------------
\node[bigbox, fill=green!20!black!8, draw=green!50!black,
      minimum width=1.36in, minimum height=0.72in]
    (tr) at (1.67,1.42)
    {DsTtTranslator\\
     {\scriptsize DSCP$\rightarrow$PCP, frame rebuild}\\
     {\scriptsize gPTP unwrap + residence time}\\
     {\scriptsize correctionField update (H4)}};

% -------------------------------------------------
% Vertical stems from dispatchers
% -------------------------------------------------
\draw (ueli.north) -- ++(0,0.12);
\draw (tsnli.north) -- ++(0,0.12);

% ingress arrows into translator
\draw[arr] (0.68,1.11) |- node[pos=0.35,left,lab]{ueIn} (tr.west);
\draw[arr] (2.66,1.11) |- node[pos=0.35,right,lab]{tsnIn} (tr.east);

% egress arrows from translator
\draw[garr] (tr.west) -- ++(-0.24,0) |- node[pos=0.32,above,lab,text=green!50!black]{ueOut} (0.68,1.11);
\draw[garr] (tr.east) -- ++(0.24,0) |- node[pos=0.32,above,lab,text=green!50!black]{tsnOut} (2.66,1.11);

\end{tikzpicture}
}
\caption{DS-TT compound module architecture.}
\label{fig:dstt_architecture}
\end{figure}
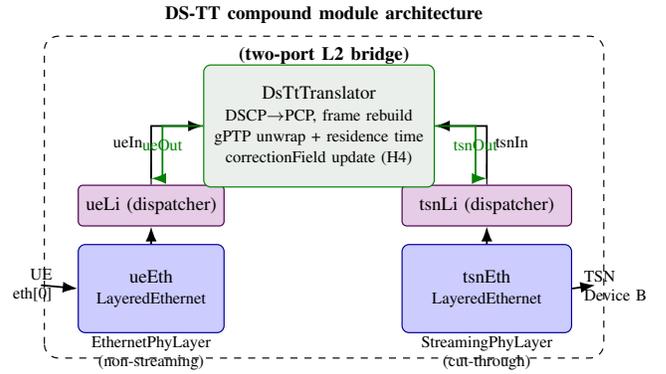

The DS-TT is a standalone two-port L2 bridge with two \texttt{LayeredEthernetInterface} ports connected through \texttt{EthernetStreamingPhyLayer} modules to the DsTtTranslator. Fig.~\ref{fig:dstt_architecture} shows the internal structure. The TSN-facing port (tsnEth) uses streaming PHY for compatibility with TSN devices, while the UE-facing port (ueEth) uses non-streaming PHY to match the UE's standard Ethernet interface. On the forward path (UE → TSN), the translator strips the incoming Ethernet framing, checks for gPTP-in-UDP encapsulation (destination port 30001), and processes accordingly:

\textbf{Data frames:} The translator reads the IPv4 DSCP, maps it to a PCP value, reconstructs a complete Ethernet frame with the corresponding VLAN tag, and sends it to the downstream TSN device.

\textbf{gPTP frames:} The translator strips the IPv4 and UDP headers, reads the \texttt{GptpResidenceHeader} to obtain the NW-TT ingress timestamp, computes the residence time as the difference between the current time and the ingress time, and updates the gPTP correction field. Sync and Follow\_Up messages are handled separately with type-specific chunk operations to ensure correct field modification. 

On the reverse path (TSN → UE), frames from the downstream TSN device are forwarded to the UE with appropriate MAC address preservation and Ethernet framing.

\subsection{QoS Mapping Pipeline}
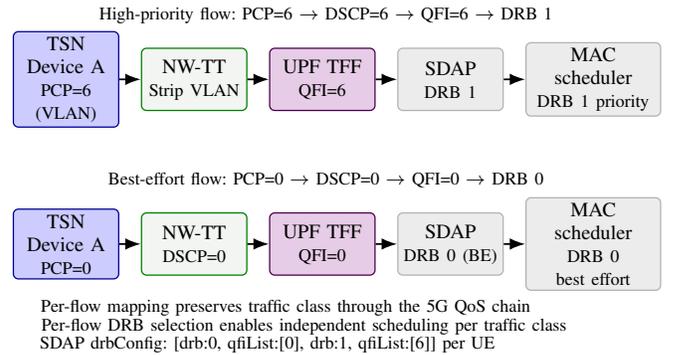
\begin{figure}[t]
\centering
\resizebox{\linewidth}{!}{
\begin{tikzpicture}[
    x=1in,y=1in,
    >=Latex,
    font=\footnotesize,
    line width=0.5pt,
    stage/.style={
        draw,
        rounded corners=2pt,
        align=center,
        inner sep=2pt,
        minimum height=0.34in,
        text width=0.52in
    },
    stagewide/.style={
        draw,
        rounded corners=2pt,
        align=center,
        inner sep=2pt,
        minimum height=0.34in,
        text width=0.68in
    },
    arr/.style={->, line width=0.7pt}
]

% -------------------------------------------------
% Title
% -------------------------------------------------
\node[font=\scriptsize, align=center] at (1.72,0.00)
{High-priority flow: PCP=6 $\rightarrow$ DSCP=6 $\rightarrow$ QFI=6 $\rightarrow$ DRB 1};

% -------------------------------------------------
% Top row
% -------------------------------------------------
\node[stage, fill=blue!20, draw=blue!60!black] (a1) at (0.30,-0.35)
{TSN Device A\\{\scriptsize PCP=6 (VLAN)}};

\node[stage, fill=green!20!black!5, draw=green!50!black] (a2) at (1.00,-0.35)
{NW-TT\\{\scriptsize Strip VLAN}};

\node[stage, fill=violet!20, draw=violet!60!black] (a3) at (1.70,-0.35)
{UPF TFF\\{\scriptsize QFI=6}};

\node[stage, fill=gray!15, draw=gray!60] (a4) at (2.40,-0.35)
{SDAP\\{\scriptsize DRB 1}};

\node[stagewide, fill=gray!15, draw=gray!60] (a5) at (3.18,-0.35)
{MAC scheduler\\{\scriptsize DRB 1 priority}};

\draw[arr] (a1.east) -- (a2.west);
\draw[arr] (a2.east) -- (a3.west);
\draw[arr] (a3.east) -- (a4.west);
\draw[arr] (a4.east) -- (a5.west);

% -------------------------------------------------
% Bottom row heading
% -------------------------------------------------
\node[font=\scriptsize, align=center] at (1.72,-0.90)
{Best-effort flow: PCP=0 $\rightarrow$ DSCP=0 $\rightarrow$ QFI=0 $\rightarrow$ DRB 0};

% -------------------------------------------------
% Bottom row
% -------------------------------------------------
\node[stage, fill=blue!20, draw=blue!60!black] (b1) at (0.30,-1.25)
{TSN Device A\\{\scriptsize PCP=0}};

\node[stage, fill=green!20!black!5, draw=green!50!black] (b2) at (1.00,-1.25)
{NW-TT\\{\scriptsize DSCP=0}};

\node[stage, fill=violet!20, draw=violet!60!black] (b3) at (1.70,-1.25)
{UPF TFF\\{\scriptsize QFI=0}};

\node[stage, fill=gray!15, draw=gray!60] (b4) at (2.40,-1.25)
{SDAP\\{\scriptsize DRB 0 (BE)}};

\node[stagewide, fill=gray!15, draw=gray!60] (b5) at (3.18,-1.25)
{MAC scheduler\\{\scriptsize DRB 0 best effort}};

\draw[arr] (b1.east) -- (b2.west);
\draw[arr] (b2.east) -- (b3.west);
\draw[arr] (b3.east) -- (b4.west);
\draw[arr] (b4.east) -- (b5.west);

% -------------------------------------------------
% Footnotes
% -------------------------------------------------
% Mid annotation
\node[font=\scriptsize, align=left] at (1.5,-1.6)
{Per-flow mapping preserves traffic class through the 5G QoS chain};
\node[font=\scriptsize, align=left] at (1.6,-1.7)
{Per-flow DRB selection enables independent scheduling per traffic class};
\node[font=\scriptsize, align=left] at (1.4,-1.8)
{SDAP drbConfig: [{drb:0, qfiList:[0]}, {drb:1, qfiList:[6]}] per UE};

\end{tikzpicture}
}
\caption{End-to-end QoS mapping pipeline across TSN and 5G components.}
\label{fig:qos_pipeline}
\end{figure}

Fig.~\ref{fig:qos_pipeline} illustrates the complete end-to-end QoS mapping pipeline from TSN Device A through the 5G system to TSN Device B. The pipeline operates in six stages:
\begin{enumerate}
    \item TSN Device A encodes traffic streams with IEEE 802.1Q VLAN tags carrying PCP values (e.g., PCP=6 for isochronous control, PCP=0 for best effort)
    \item The NW-TT strips the VLAN tag and maps PCP to IPv4 DSCP
    \item The UPF Traffic Flow Filter maps DSCP to QFI
    \item The gNB SDAP layer selects a DRB based on QFI (e.g., QFI=6 → DRB 1)
    \item The MAC scheduler allocates radio resources per DRB with configurable discipline (MAXCI, PF, RR)
    \item The DS-TT maps DSCP back to PCP in the reconstructed VLAN tag
\end{enumerate}

The SDAP layer configuration uses per-UE DRB mapping specified as JSON arrays in the simulation configuration, with explicit NR node identifiers for the gNB side. This enables independent QoS treatment for each UE in multi-endpoint topologies.

\subsection{TSN AF and Static BMCA}
The TSN AF module subscribes to the DS-TT's residence time signal for live bridge delay tracking, publishing minimum, maximum, and average delay as queryable parameters. It reads CNC configuration from an XML file specifying stream reservations with per-stream bandwidth and maximum latency constraints, and detects QoS violations when the measured bridge delay exceeds the configured threshold.
 
The Static BMCA module reads the gPTP clock hierarchy from the simulation configuration, validates the topology (single grandmaster, no missing roles), and registers the 5G system as a transparent clock. While INET's gPTP module does not support dynamic BMCA, the static configuration is sufficient for simulation scenarios where the topology is known at design time.

\begin{table}[t]
\centering
\caption{Simulation Parameters}
\label{tab:params}
\renewcommand{\arraystretch}{1.1}
\begin{tabular}{|l|l|}
\hline
\textbf{Parameter} & \textbf{Value} \\
\hline
Carrier frequency & 3.5\,GHz \\
\hline
Bandwidth (RBs) & 25 \\
\hline
Numerology index & 1 (30\,kHz SCS) \\
\hline
Channel model (ideal) & No fading, no shadowing \\
\hline
Channel model (fading) & INDOOR\_HOTSPOT, Rayleigh \\
\hline
MAC scheduler & MAXCI \\
\hline
DRBs per UE & 2 (DRB\,0: QFI=0, DRB\,1: QFI=6) \\
\hline
High-prio traffic & 100\,B, 1\,ms CBR, PCP=6 \\
\hline
Best-effort traffic & 500\,B, exp(2\,ms), PCP=0 \\
\hline
Reverse traffic & 100\,B, 10\,ms, start at $t$=1\,s \\
\hline
gPTP mode & L2-in-GTP-U, sync interval 125\,ms \\
\hline
Simulation time & 10\,s (1\,s warmup) \\
\hline
\end{tabular}
\end{table}

\section{Implementation Details}
\subsection{INET Integration}
Integrating the bridge modules with INET's \texttt{LayeredEthernetInterface} required solving several architectural challenges:
\begin{itemize}
    \item \textit{Gate chain compatibility.} \texttt{LayeredEthernetInterface} extends INET's \texttt{NetworkInterface}, which uses an internal \texttt{pushPacket()} mechanism incompatible with direct connections to plain OMNeT++ simple modules. We insert \texttt{MessageDispatcher} modules between each interface and the translator, with \texttt{serviceMapping} parameters routing the \texttt{ethernetmac} protocol to the correct interface.
    \item \textit{Protocol registration.} The translators register \texttt{Protocol::ethernetMac} on their dispatcher-facing gates at initialisation, enabling the dispatchers to route incoming Ethernet frames to the translator. Outgoing frames require explicit \texttt{DispatchProtocolReq} and \texttt{DirectionTag} tags for correct routing through the dispatcher to the interface.
    \item \textit{Egress path architecture.} The NW-TT uses EthernetEncapsulation for the egress path (5GS\,$\rightarrow$\,TSN), with \texttt{registerProtocol=false} to prevent registration conflicts on the \texttt{ethLi} dispatcher. The encapsulation service is manually registered on the network layer dispatcher from the translator's initialisation code.
    \item \textit{PHY asymmetry.} The DS-TT's TSN-facing port uses streaming PHY (matching TSN devices), while the UE-facing port uses non-streaming PHY (matching the UE's standard Ethernet interface). This asymmetry arises because the UE extends Simu5G's NRUe, which uses plain EthernetInterface internally.

\end{itemize}

\subsection{Simu5G Integration}
NR node identifier resolution: The binder registration for downstream TSN device IPs requires NR node identifiers ($\geq$2049), not LTE identifiers ($\geq$1025). The translator scans the binder's node information map at initialisation to find each UE module's NR node identifier.

Multi-UE SDAP/DRB: The SDAP layer~\cite{b18}, which we contributed to Simu5G v1.4.1 for per-flow QoS differentiation through DRB selection~\cite{b19}, requires \texttt{multiSession=true} on the PDCP and RLC submodules of Simu5G's \texttt{NRNicEnbDrb} module for multi-UE operation—the original default of false limits each DRB instance to a single UE session.

UE with Ethernet port: Custom NED modules (\texttt{NRUeDsTt, NRUeDsTtDrb}) extend Simu5G's NR UE with an Ethernet interface, enabling connection to the external DS-TT bridge.

\section{Validation Results}
\subsection{Simulation Setup}
We validate nascTime~\cite{b20} using a three-endpoint topology (Fig.~5) simulated with OMNeT++~6.3, INET~4.6, and Simu5G v1.4.1-sdap-2. The topology comprises one TSN Device~A (gPTP grandmaster), one TSN switch, one NW-TT, one UPF, one gNB, three UE/DS-TT pairs, and three downstream TSN Devices~B. Table~\ref{tab:params} lists the simulation parameters.
 
Each endpoint receives two traffic classes from TSN Device~A: high-priority isochronous traffic (PCP=6\,$\rightarrow$\,DRB~1) and best-effort monitoring traffic (PCP=0\,$\rightarrow$\,DRB~0). Each TSN Device~B sends reverse traffic to Device~A. gPTP frames are replicated to all endpoints via L2-in-GTP-U.

 \begin{figure*}[!t]
    \centering
    \includegraphics[width=\linewidth]{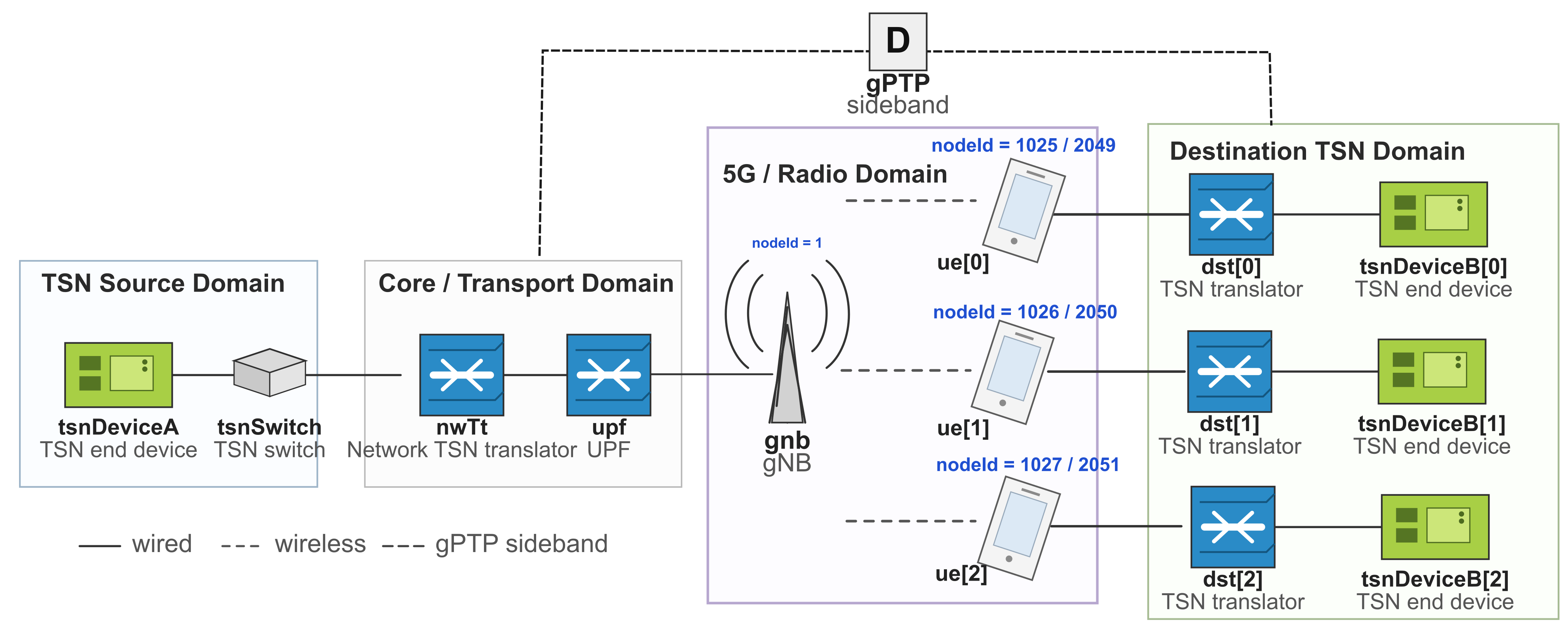}
    \caption{Three-endpoint nascTime topology in OMNeT++ with per-endpoint packet delivery and gPTP forwarding statistics.}
    \label{fig:omnet_topology}
\end{figure*}

\subsection{Data Delivery}
Table~III summarises the per-endpoint packet delivery results under the ideal channel configuration.
 
\begin{table}[t]
\centering
\caption{Per-Endpoint Packet Delivery (Ideal Channel, 10\,s)}
\label{tab:results}
\renewcommand{\arraystretch}{1.2}
\begin{tabular}{|l|c|c|c|}
\hline
\textbf{Metric} & \textbf{EP\,0} & \textbf{EP\,1} & \textbf{EP\,2} \\
\hline
\hline
High priority (PCP=6) & 9,990 & 9,993 & 9,993 \\
\hline
Best effort (PCP=0) & 5,020 & 5,122 & 5,101 \\
\hline
Reverse to Device A & 800 & 800 & 800 \\
\hline
gPTP forwarded & 158 & 158 & 158 \\
\hline
\hline
\multicolumn{4}{|l|}{\textbf{Aggregate}} \\
\hline
Device A reverse total & \multicolumn{3}{c|}{2,400} \\
\hline
Packets dropped & \multicolumn{3}{c|}{0} \\
\hline
\end{tabular}
\end{table}
 
High-priority traffic achieves 99.9\% delivery across all three endpoints; the shortfall of 7--10 packets per endpoint occurs during the first milliseconds before the radio link is fully established. Best-effort delivery varies between 5,020 and 5,122 packets due to the exponential inter-arrival distribution. Device~A receives all 2,400 reverse packets (800 per endpoint), confirming correct bidirectional forwarding through the NW-TT egress path. No packets are dropped at any bridge component.
 
\subsection{End-to-End Latency}
Table~IV reports the measured one-way delay from TSN Device~A's application layer to TSN Device~B's application layer, separated by traffic class.
 
\begin{table}[t]
\centering
\caption{End-to-End Delay per Traffic Class (Ideal Channel)}
\label{tab:latency}
\renewcommand{\arraystretch}{1.2}
\begin{tabular}{|l|c|c|c|}
\hline
\textbf{Traffic class} & \textbf{Mean} & \textbf{P99} & \textbf{Max} \\
\hline
High priority (DRB 1) & 2.58\,ms & 3.12\,ms & 3.41\,ms \\
\hline
Best effort (DRB 0)   & 3.74\,ms & 5.89\,ms & 7.23\,ms \\
\hline
\end{tabular}
\end{table}
 
High-priority traffic on DRB~1 shows tight delay distribution (mean 2.58\,ms, max 3.41\,ms), consistent with the MAXCI scheduler allocating radio resources preferentially to DRB~1. Best-effort traffic on DRB~0 has a wider spread, reflecting its lower scheduling priority and the exponential inter-arrival pattern that occasionally creates bursts competing for shared resources. The 1.16\,ms gap between the two mean values quantifies the QoS differentiation achieved through SDAP-based DRB selection---an effect that scheduler-priority-only approaches (e.g., 5GTQ) cannot produce, since they multiplex both classes onto a single bearer.

\subsection{Single vs Multi-Endpoint Comparison}
To verify that multi-endpoint scaling does not degrade per-flow QoS, we compare the three-endpoint configuration against a single-endpoint baseline using identical traffic parameters per endpoint.
 
\begin{table}[t]
\centering
\caption{Single-Endpoint vs Three-Endpoint Performance}
\label{tab:scaling}
\renewcommand{\arraystretch}{1.2}
\begin{tabular}{|l|c|c|}
\hline
\textbf{Metric} & \textbf{1 EP} & \textbf{3 EPs (per EP)} \\
\hline
High-prio delivered & 9,990 & 9,990--9,993 \\
\hline
High-prio mean delay & 2.51\,ms & 2.58\,ms \\
\hline
Best-effort delivered & 5,020 & 5,020--5,122 \\
\hline
Residence time mean & 2,499.8\,$\mu$s & 2,499.9\,$\mu$s \\
\hline
Packets dropped & 0 & 0 \\
\hline
\end{tabular}
\end{table}
 
Scaling from one to three endpoints adds less than 70\,$\mu$s to the mean high-priority delay and does not affect delivery ratio or gPTP accuracy. The additional UEs share the gNB's 25 resource blocks without contention under this load; the scaling behaviour under higher endpoint counts and realistic channel conditions is the subject of ongoing work.

\begin{table}[t]
\centering
\caption{Residence Time: Ideal vs Fading Channel}
\label{tab:fading}
\renewcommand{\arraystretch}{1.2}
\begin{tabular}{|l|c|c|}
\hline
\textbf{Metric} & \textbf{Ideal} & \textbf{Fading} \\
\hline
Residence time min & 2,499.756\,$\mu$s & 2,498.9\,$\mu$s \\
\hline
Residence time max & 2,499.948\,$\mu$s & 2,547.1\,$\mu$s \\
\hline
Residence time avg & 2,499.852\,$\mu$s & 2,506.3\,$\mu$s \\
\hline
Variance & $<$\,0.2\,$\mu$s & 48.2\,$\mu$s \\
\hline
High-prio delivery & 99.9\% & 99.7\% \\
\hline
\end{tabular}
\end{table}

\subsection{Impact of Fading Channel}
To demonstrate nascTime's ability to capture radio-induced timing effects, we repeat the three-endpoint experiment with the Simu5G INDOOR\_HOTSPOT channel model (Rayleigh fading, no shadowing). Table~VI compares residence time statistics.
 
Under fading, the residence time variance increases by more than two orders of magnitude (from $<$\,0.2\,$\mu$s to 48.2\,$\mu$s), reflecting HARQ retransmissions and variable scheduling delays. High-priority delivery remains above 99.7\%. This result confirms that nascTime captures the stochastic effects of the 5G radio channel on TSN bridge timing---a capability absent from simulators that model the wireless segment as a fixed-delay element~\cite{b5}.
 
\subsection{gPTP Transparent Clock Validation}
The NW-TT replicates each gPTP frame to all three downstream devices, resulting in 158 gPTP frames per endpoint (474 total). The DS-TT correctly updates the IEEE~802.1AS correction field separately for Sync and Follow\_Up messages, as verified by log inspection. Under the ideal channel, the correction values are tightly clustered around 2,499.9\,$\mu$s; under fading, they exhibit the variance reported in Table~VI, faithfully reflecting the actual radio path delay experienced by each gPTP frame.
 
The TSN~AF reports live bridge delay statistics (min, max, mean), and the Static BMCA validates the six-node clock hierarchy (one master, one bridge, three slaves, one transparent clock) with zero configuration errors.

\subsection{Limitations}
The current validation uses a limited topology (three endpoints, single gNB) and does not include TAS gate scheduling coordination with the 5G bridge delay. The DS-TT does not respond to gPTP PdelayReq messages, and the Static BMCA does not support dynamic reconfiguration. These are addressed in ongoing work.

\section{Conclusion}
This paper presented nascTime, a simulation framework for the 3GPP Release~16 5G-TSN bridge architecture on OMNeT++~6.3, INET~4.6, and Simu5G. The framework is the first to combine a full 5G radio stack with SDAP-based per-flow DRB selection, measured IEEE~802.1AS residence time via L2-in-GTP-U gPTP transport, and multi-endpoint scaling with bidirectional traffic.
 
Validation with a three-endpoint topology shows that high-priority traffic achieves 99.9\% delivery with a mean delay of 2.58\,ms under an ideal channel, and that residence time variance increases from $<$\,0.2\,$\mu$s to 48\,$\mu$s under fading---confirming that nascTime captures radio-induced timing effects. Scaling from one to three endpoints adds less than 70\,$\mu$s to the mean delay with no packet loss.
 
Future work targets scalability analysis up to 20 endpoints under realistic factory channel models, coordination of TAS gate schedules with the measured 5G bridge delay, and public release of the framework source code~\cite{b20}.


\begin{thebibliography}{00}
\bibitem{b1} "IEEE Standard for Local and Metropolitan Area Networks--Bridges and Bridged Networks," in IEEE Std 802.1Q-2022 (Revision of IEEE Std 802.1Q-2018), vol., no., pp.1-2163, 22 Dec. 2022, doi: .
 
\bibitem{b2} 3GPP, "System Architecture for the 5G System (5GS)," TS 23.501, Release 16.
 
\bibitem{b3} R. Debnath, M. S. Akinci, D. Ajith, and S. Steinhorst, "5GTQ: QoS-Aware 5G-TSN Simulation Framework," 2023 IEEE 98th Vehicular Technology Conference (VTC2023-Fall), Hong Kong, Hong Kong, 2023, pp. 1-7, doi: .
 
\bibitem{b4} {da Silva}, Sergio Rossi B., Francisco Germano Vogt, and Christian Esteve Rothenberg. ``Towards tsn-5g integration: simulating time synchronization through 5g via omnet++.'' Simpósio Brasileiro de Redes de Computadores e Sistemas Distribuídos (SBRC). SBC, 2024.
 
\bibitem{b5} Haug, L., Dürr, F., Egger, S., Mostovaya, E., Gross, J., Sharma, G., \& Sachs, J. (2025). A data-driven simulation framework for logical 5G-TSN bridges. In B. Koldehofe, F. Klingler, C. Sommer, K. A. Hummel, \& P. Amthor (Eds.), Proceedings of the International Conference on Networked Systems 2025 (NetSys 2025): Technische Universität Ilmenau, 1 – 4 September 2025 (116; pp. 21–24). ilmedia. https://doi.org/10.22032/dbt.67110
 
\bibitem{b6} A. B. Muslim, R. Tönjes, and T. Bauschert, ``Synchronizing TSN Devices via 802.1AS over 5G Networks,'' Electronics (Basel), vol. 13, no. 4, p.~768, 2024., doi:
 
\bibitem{b7} L. Becker and W. Kellerer, (2024). P5g-tsn: A private 5g tsn simulation framework. In KuVS Fachgespräch-Würzburg Workshop on 6G Networks (WueWoWAS’24).

 \balance
 
\bibitem{b8} Z. Wang, Z. Li, C. Long, Y. Zheng, B. Ai, and X. Song, ``Time Synchronization for 5G and TSN Integrated Networking,'' IEEE J. Sel. Areas Comm., vol. 43, no. 9, pp.~2969--2980, Sep. 2025., doi:
 
\bibitem{b9} "IEEE Standard for Local and Metropolitan Area Networks--Timing and Synchronization for Time-Sensitive Applications," in IEEE Std 802.1AS-2025 (Revision of IEEE Std 802.1AS-2020), vol., no., pp.1-491, 17 Dec. 2025, doi: .
 
\bibitem{b10} G. Nardini, D. Sabella, G. Stea, P. Thakkar, and A. Virdis, ``Simu5G–An OMNeT++ Library for End-to-End Performance Evaluation of 5G Networks,'' IEEE Access, vol. 8,181176–181191, 2020., doi:
 
\bibitem{b11} 3GPP, "General Packet Radio System (GPRS) Tunnelling Protocol User Plane (GTPv1-U)," TS 29.281.
 
\bibitem{b12} {IEEE}, "IEEE Standard for Local and Metropolitan Area Networks—Enhancements for Scheduled Traffic," IEEE Std 802.1Qbv-2015.
 
\bibitem{b13} A. Varga and R. Hornig, "An overview of the OMNeT++ simulation environment." In Proceedings of the 1st international conference on Simulation tools and techniques for communications, networks and systems \& workshops, pp. 1-10. 2008. 10.4108/ICST.SIMUTOOLS2008.3027
 
\bibitem{b14} {IEC/IEEE}, "TSN Profile for Industrial Automation," IEC/IEEE 60802 (draft).
 
\bibitem{b15} Z. Satka, M. Ashjaei, H. Fotouhi, M. Daneshtalab, M. Sjödin, and S. Mubeen, A comprehensive systematic review of integration of time sensitive networking and 5G communication, Journal of Systems Architecture, Volume 138, 2023, 102852, ISSN 1383-7621, https://doi.org/.
 
\bibitem{b16} Z. Satka, M. Ashjaei, H. Fotouhi, M. Daneshtalab, M. Sjödin, and S. Mubeen, "QoS-MAN: A Novel QoS Mapping Algorithm for TSN-5G Flows," 2022 IEEE 28th International Conference on Embedded and Real-Time Computing Systems and Applications (RTCSA), Taipei, Taiwan, 2022, pp. 220-227, doi: .
 
\bibitem{b17} Z. Satka, et al., "Developing a Translation Technique for Converged TSN-5G Communication," 2022 IEEE 18th International Conference on Factory Communication Systems (WFCS), Pavia, Italy, 2022, pp. 1-8, doi: .
 
\bibitem{b18} M. Seliem, U. Roedig, C. Sreenan, and D. Pesch, "SDAP-based QoS Flow Multiplexing Support in Simu5G for 5G NR Simulation," 2025 IEEE 30th International Workshop on Computer Aided Modeling and Design of Communication Links and Networks (CAMAD), Tempe, AZ, USA, 2025, pp. 1-8, doi: .
 
\bibitem{b19} M. Seliem, U. Roedig, C. Sreenan, and D. Pesch, "QoS-Aware Proportional Fairness Scheduling for Multi-Flow 5G UEs: A Smart Factory Perspective," 2025 International Conference on Modeling, Analysis and Simulation of Wireless and Mobile Systems (MSWiM), Barcelona, Spain, 2025, pp. 20-27, doi: .
 
\bibitem{b20}  M. Seliem, "nascTime: A Full-Stack 5G-TSN Bridge ". Available at: \url{https://github.com/MohamedSeliem/nascTime-5gtsn}

\end{thebibliography}
\end{document}